\documentclass[aps,prl,reprint,preprintnumbers,showpacs]{revtex4-1}

\usepackage[rgb]{xcolor}
\usepackage{color}
\usepackage{amssymb,amsmath,mathrsfs,graphicx,bbm,dsfont,booktabs,mathtools,braket,lmodern,tikz}

\DeclareMathAlphabet{\mathpzc}{OT1}{pzc}{m}{it}

\DeclareMathAlphabet{\mathsc}{T1}{lmr}{m}{scsl}

\newcommand{\bea}{\begin{eqnarray}}
\newcommand{\eea}{\end{eqnarray}}
\def\be{\begin{equation}}
\def\ee{\end{equation}}
\newcommand{\bei}{\begin{itemize}}
\newcommand{\eei}{\end{itemize}}
\newcommand{\bee}{\begin{enumerate}}
\newcommand{\eee}{\end{enumerate}}

\def\vk{\varkappa}

\def\z{\zeta}
\def\eps{\epsilon}

\def\pa {\partial}

\def\ads{{\rm AdS}_5\times {\rm S}^5}

\def\ads{{\rm AdS}_5\times {\rm S}^5}

\def\am{{\rm am}}
\def\am0{{\rm am}_0}

\expandafter\def\expandafter\bfseries\expandafter{\bfseries\ifmmode\else\boldmath\fi}
\expandafter\def\expandafter\mdseries\expandafter{\mdseries\ifmmode\else\unboldmath\fi}
\expandafter\def\expandafter\normalfont\expandafter{\normalfont\ifmmode\else\unboldmath\fi}

\definecolor{grey}{rgb}{0.4,0.4,0.5}
\definecolor{darkgreen}{rgb}{0,0.5,0}
\definecolor{darkred}{rgb}{0.6,0.0,0}
\definecolor{lightbrown}{rgb}{1,0.9,0.8}
\definecolor{brown}{rgb}{0.6,0.3,0.3}
\definecolor{darkblue}{rgb}{0,0,0.8}
\definecolor{darkmagenta}{rgb}{0.5,0,0.5}

\begin{document}

\title{The $\ads$ mirror model as a string sigma model}
\author{Gleb Arutyunov}
\email{g.e.arutyunov@uu.nl}
\thanks{Correspondent fellow at Steklov Mathematical Institute, Moscow.}
\affiliation{Institute for Theoretical Physics and Spinoza Institute, Utrecht University, Leuvenlaan 4, 3584 CE Utrecht, The Netherlands}
\author{Stijn J. van Tongeren}
\email{svantongeren@physik.hu-berlin.de}
\affiliation{Institut f\"ur Mathematik und Institut f\"ur Physik, Humboldt-Universit\"at zu Berlin, IRIS Geb\"aude, Zum Grossen Windkanal 6, 12489 Berlin}

\begin{abstract}
Doing a double Wick rotation in the worldsheet theory of the light cone ${\rm AdS}_5\times {\rm S}^5$ superstring results in an inequivalent, so-called mirror theory that plays a central role in the field of integrability in AdS/CFT. We show that this mirror theory can be interpreted as the light cone theory of a free string on a different background. This background is related to $\mathrm{dS}_5 \times \mathrm{H}^5$ by a double T duality, and has hidden supersymmetry. The geometry can also be extracted from an integrable deformation of the ${\rm AdS}_5\times {\rm S}^5$ sigma model, and we prove the observed mirror duality of these deformed models at the bosonic level as a byproduct. While we focus on ${\rm AdS}_5\times {\rm S}^5$, our results apply more generally.
\end{abstract}


\preprint{HU-EP-14/21, HU-MATH-14/12}
\preprint{ITP-UU-14/18, SPIN-14/16}

\maketitle

Integrability has and continues to be of central importance in furthering our understanding of the  AdS/CFT correspondence \cite{Maldacena:1997re}, giving important insights into finite coupling quantum field theory. Through fruitful interplay between results on both sides of the correspondence, remarkable progress has been made in the spectral problem in particular \footnote{For reviews see \cite{Arutyunov:2009ga,Beisert:2010jr}.}. Namely, we can describe scaling dimensions in planar $\mathcal{N}=4$ supersymmetric Yang-Mills theory (SYM) at finite 't Hooft coupling through the corresponding energy levels of a string on $\ads$, and these energy levels can be computed exactly thanks to the integrability of the string \cite{Bena:2003wd}. More precisely, under the assumption that integrability persists at the quantum level, the spectrum of the $\ads$ superstring can be determined by means of the thermodynamic Bethe ansatz applied to a doubly Wick rotated version of its worldsheet theory \cite{Zamolodchikov:1989cf,Dorey:1996re}, as put forward in \cite{Ambjorn:2005wa} and worked out in \cite{Arutyunov:2007tc,Arutyunov:2009zu,Arutyunov:2009ur,Bombardelli:2009ns,Gromov:2009bc}. Since the light cone gauge fixed $\ads$ string is not Lorentz invariant however, this double Wick rotation results in an inequivalent quantum field theory, the so-called mirror theory \cite{Arutyunov:2007tc}. The associated mirror transformation also appears extensively in the exact description of polygonal Wilson loops or equivalently planar scattering amplitudes \cite{Basso:2013vsa,Basso:2013aha,Basso:2014koa}. 
Given the central importance of the mirror theory, we would like to elevate it beyond the status of a technical tool. This raises a question that has gone unanswered since the model's introduction in 2007, namely whether the mirror theory itself can arise directly by light cone gauge fixing a free string on some background. Here we show that this is the case, giving an interesting relation between strings on different backgrounds different from other well known dualities. Our results extend to the integrable deformation of the $\ads$ superstring of \cite{Delduc:2013qra}, show an intriguing link to de Sitter space, and indicate new backgrounds with hidden supersymmetry in the associated string sigma model.

Our construction is based on a simple observation regarding the bosonic light cone string action that directly produces the desired `mirror' metric for a fairly generic class of backgrounds. The doubly Wick-rotated light cone action is obtained by a formal exchange and inversion of the metric components of the two directions making up the light cone coordinates, and a sign flip on the $B$ field. It is not obvious that the resulting metric is part of a string background, but we demonstrate explicitly that the mirror version of $\ads$ is a solution of type IIB supergravity with nontrivial dilaton and Ramond-Ramond five-form. The double Wick rotation of the corresponding Green-Schwarz fermions is compatible with our considerations, as we have explicitly verified at quadratic level and will report on elsewhere \cite{Arutyunov:2014jfa}.

Interestingly, the mirror space we obtain from $\ads$ is formally related to $\mathrm{dS}_5 \times \mathrm{H}^5$ by a double T duality, $\mathrm{H}^5$ being the five-dimensional hyperboloid. The space also has a curvature singularity, but this is not necessarily problematic for the string sigma model. In fact, the mirror sigma model inherits the symmetries and in particular the integrability of the light cone $\ads$ sigma model, hinting at well defined behavior. The bosonic $\mathfrak{su}(2)^{\oplus4} \subset \mathfrak{psu}(2|2)^{\oplus2} \oplus \mathbb{H}$ symmetry of the  $\ads$ light cone string matches the $\mathfrak{so}(4)^{\oplus2}$ symmetry of our background ($\mathbb{H}$ is a central element corresponding to the worldsheet Hamiltonian). The supersymmetry of the mirror model is not realized through superisometries however, as the mirror background admits no Killing spinors. This is natural because the central element of the symmetry algebra of the mirror theory is nonlinearly related to its Hamiltonian ($\mathbb{C} \sim \sinh\frac{\tilde{\mathbb{H}}}{2}$) \cite{Arutyunov:2007tc}, so that the full superalgebra should be nonlinearly realized on the fermions of the mirror background.

The mirror sigma model's integrability is not obvious from its geometry. Interestingly however, we can obtain (the bosonic part of) this sigma model as a limit of the integrable deformation of the $\ads$ coset sigma model constructed in \cite{Delduc:2013qra}. The way to take this limit naturally follows when we use the main observation of this letter to prove the mirror duality \cite{Arutynov:2014ota} of these deformed models (at the bosonic level).
Provided technical complications in extracting the fermions in these deformed models can be overcome, this relation would manifest classically integrability and $\kappa$ symmetry of the mirror background.

\section{Double Wick rotations of light cone gauge fixed strings}
\label{sec:LCGfixedstring}

We want to understand whether the double Wick rotated light cone worldsheet theory of a string on a given background can be realized by light cone gauge fixing a string on another background. At the bosonic level this works quite elegantly. We will consider $d$-dimensional backgrounds with coordinates $\{t,\phi,x^\mu\}$ and metric
\begin{equation}
\nonumber
ds^2 \equiv g_{\mathsc{mn}} \, d x^\mathsc{m} d x^\mathsc{n} =-g_{tt} dt^2+g_{\phi\phi} d\phi^2 + g_{\mu\nu}dx^\mu dx^\nu,
\end{equation}
the components depending only on the transverse coordinates $x^\mu$, and $B$ fields that are nonzero only in the transverse directions. The bosonic string action is given by
\begin{equation}
\label{eq:stringaction}\nonumber
S=-\tfrac{T}{2}\int{\rm d}\tau{\rm d}\sigma \, \left(g_{\mathsc{mn}}\, dx^\mathsc{m} dx^\mathsc{n} - B_{\mathsc{mn}}\, dx^\mathsc{m} \hspace{-3pt} \wedge dx^\mathsc{n}\right) ,
\end{equation}
where $T$ is the string tension. To fix a light cone gauge we introduce light cone coordinates \footnote{Strictly speaking we are fixing a particular form of the generalized uniform light cone gauge \cite{Arutyunov:2006gs}, also known as the temporal gauge. Our discussion also directly applies to the general form.}
\begin{equation}
\nonumber
x^+=t,\hspace{20pt}x^{-}=\phi-t ,
\end{equation}
and fix (in the first order formalism)
\begin{equation}
\nonumber
x^+ = \tau, \hspace{20pt} p^+ = 1.
\end{equation}
The action then takes the form (see e.g. \cite{Arutynov:2014ota})
\begin{equation}
\nonumber
S= T \int{\rm d}\tau{\rm d}\sigma \, \left(1 -\sqrt{Y}  + B_{\mu\nu}\dot{x}^{\mu}x'^{\nu}\right),
\end{equation}
where
\begin{align}
\nonumber
Y=&\,(\dot{x}_{\mu}x'^{\mu})^2 -(\dot{x}_{\mu}\dot{x}^{\mu}-g_{tt})(x'_{\nu}x'^{\nu}+1/g_{\phi\phi}) ,
\end{align}
and dots and primes refer to temporal and spatial derivatives on the worldsheet, having rescaled $\sigma\rightarrow T \sigma$. We now observe that a double Wick rotation of the worldsheet coordinates
\begin{equation}
\nonumber
\tau  \rightarrow i \tilde{\sigma} ,\hspace{20pt} \sigma \rightarrow - i \tilde{\tau}.
\end{equation}
gives an action of the same form, with $g_{tt}$ interchanged for $1/g_{\phi\phi}$ and $B$ for $-B$. Thus we can obtain the double Wick rotated worldsheet theory also directly by gauge fixing a string on a background with a metric with $g_{tt}$ and $1/g_{\phi\phi}$ interchanged and a $B$ field with opposite sign. Note that we generically denote quantities in the double Wick rotated theory (the mirror theory) by tildes.

Taking this construction and feeding it the metric of $\ads$ in global coordinates \footnote{Naturally our conventions are precisely such that $t$ and $\phi$ are the coordinates used in the light cone gauge fixing that leads up to the $\ads$ mirror model.}
\begin{align}\nonumber
ds^2 =&\,-(1+\rho^2)dt^2 + \frac{d\rho^2}{1+\rho^2} + \rho^2 d\Omega_{3}\\
 &\,\hspace{12pt}+ (1-r^2) d\phi^2 + \frac{dr^2}{1-r^2} +r^2 d\Omega_{3} ,\nonumber
\end{align}
we directly obtain
\begin{align}
\nonumber
ds^2 =&\,\frac{1}{1-r^2}(-dt^2 + dr^2) + r^2 d\Omega_{3} \\
&\, \hspace{6pt}+\frac{1}{1+\rho^2}\left(d\phi^2  + d\rho^2\right) + \rho^2 d\Omega_{3} ,\label{eq:mirrormetric}
\end{align}
the metric that would result in the (bosonic) mirror theory. The transverse directions should not be affected by this transformation, but for the light cone directions this is more subtle and the $\phi$ direction need not keep its range. For the mirror version of $\ads$ we will take $\phi$ noncompact in fact, upon considering the relation of our mirror space to the spaces appearing in the deformed sigma models of \cite{Delduc:2013qra}.

\section{Deformed $\mathrm{AdS}_5 \times \mathrm{S}^5$ coset sigma models}
\label{sec:deformedstrings}

The family of deformed sigma models of \cite{Delduc:2013qra} can be labeled by a parameter $\varkappa \in [0,\infty)$, the undeformed $\ads$ string sigma model sitting at $\varkappa = 0$. The corresponding metric is given by \cite{Arutyunov:2013ega}
\begin{align}
\nonumber
ds^2 = \,&
-\frac{f_+(\rho)}{f_-(\varkappa \rho)} dt^2 +\frac{1}{f_+(\rho)f_-(\varkappa \rho)} d\rho^2 + \rho^2 d\Theta^\rho_{3}\\
&
 + \frac{f_-(r)}{f_+(\varkappa r)} d\phi^2 +\frac{1}{f_-(r)f_+(\varkappa r)} dr^2 +r^2 d\Theta^r_{3},\nonumber
\end{align}
where $f_{\pm}(x) = 1\pm x^2$ and $d\Theta_{3}$ is a deformation of the three-sphere metric in Hopf coordinates
\begin{align}
d\Theta^\rho_{3} \equiv & \, \frac{1}{1+\varkappa^2 \rho^4 \sin^2 \zeta} (d\zeta^2 + \cos^2 \zeta d\psi_1^2) +  \sin^2 \zeta d \psi_2^2,\nonumber
\\
d\Theta^r_{3} \equiv & \, \frac{1}{1+\varkappa^2 r^4 \sin^2 \xi} (d\xi^2 + \cos^2 \xi d\chi_1^2) +  \sin^2 \xi d \chi_2^2.\nonumber
\end{align}
The $B$ field is given by
\begin{align}
B = &\, \varkappa \Big(\frac{\rho ^4 \sin 2 \zeta}{1+ \varkappa ^2 \rho ^4 \sin ^2\z} d\psi_1 \wedge d\z \nonumber\\
& \hspace{1.2cm} -\frac{ r^4 \sin 2 \xi }{1+ \varkappa ^2 r^4 \sin^2\xi} d\chi_1 \wedge d\xi\Big),\nonumber
\end{align}
vanishing at $\vk=0$. The tension can be conveniently parameterized as
\begin{equation}
\nonumber
T = g \sqrt{1+\varkappa^2}.
\end{equation}
The range of $\rho$ is restricted to $[0,1/\varkappa)$ to preserve the timelike nature of $t$, with a curvature singularity at $\rho=1/\varkappa$. At $\varkappa=0$ there is no singularity but rather the conformal boundary of anti-de Sitter space at $\rho=\infty$.

Now let us introduce rescaled coordinates
\begin{align}
\nonumber
\tilde{t} & = \varkappa t,  & \tilde{\phi} & = \varkappa \phi, & \tilde{r} & = \varkappa \rho,  & \tilde{\rho} & = \varkappa r,
\end{align}
and relabeled coordinates
\begin{align}
\nonumber
\tilde{\xi} & = \z,  & \tilde{\z} & = \xi, &
\tilde{\chi}_{i} & = \psi_{i},  & \tilde{\psi}_{i} & = \chi_{i}.
\end{align}
If we then also introduce $\tilde{\vk}= 1/\vk$,
the metric $ds^2 = \tilde{\vk}^2 \tilde{ds}^{\raisebox{-2.5pt}{\scriptsize 2}}$ becomes
\begin{align}
\nonumber
\tilde{ds}^{\raisebox{-2.5pt}{\scriptsize 2}} = \,&-\frac{f_+(\tilde{\vk} \tilde{r})}{f_-(\tilde{r})} d\tilde{t}^2 +\frac{1}{f_+(\tilde{\vk} \tilde{r})f_-(\tilde{r})} d\tilde{r}^2 + \tilde{r}^2 d\Theta^{\tilde{r}}_{3}\\
&
 + \frac{f_-(\tilde{\vk} \tilde{\rho})}{f_+(\tilde{\rho})} d\tilde{\phi}^2 +\frac{1}{f_-(\tilde{\vk} \tilde{\rho})f_+(\tilde{\rho})} d\tilde{\rho}^2 +\tilde{\rho}^2 d\Theta^{\tilde{\rho}}_{3},\nonumber
\end{align}
where now the $d \Theta_3$ factors contain tildes on $\varkappa$ and the angles. Up to the tildes and the factor of $\tilde{\vk}^{2}$ this is nothing but the deformed metric we started with, with $g_{tt}$ and $1/g_{\phi\phi}$ interchanged. Similarly, the $B$ field precisely picks up a sign in addition to tildes and a factor of $\tilde{\vk}^{2}$. This factor can now be absorbed in the string tension.

Identifying quantities with and without tildes puts us squarely in the situation of the previous section, proving that the deformed bosonic light cone theory at tension $T(g)$ and deformation value $\varkappa$ is equal to the double Wick rotated theory at tension $T(\tilde{\varkappa} g)$ and deformation value $\tilde{\vk}$. This is precisely the mirror duality observed in \cite{Arutynov:2014ota}, which we have now proven at the bosonic level.

Just as the undeformed limit $\vk\rightarrow0$ gives the metric of $\ads$ , the maximally deformed limit $\tilde{\vk}\rightarrow0$ gives precisely the mirror metric \eqref{eq:mirrormetric}. From this perspective the mirror $\phi$ coordinate is naturally noncompact.  Let us now discuss the mirror space in more detail.

\section{The $\mathrm{AdS}_5 \times \mathrm{S}^5$ mirror background}
\label{sec:adsmirrorbackground}

We obtained the mirror space \eqref{eq:mirrormetric} directly from $\ads$, but it is also closely related to $\mathrm{dS}_5 \times \mathrm{H}^5$. Namely, (by construction) our mirror space turns into $\mathrm{dS}_5 \times \mathrm{H}^5$ upon applying a timelike T duality in $t$ and a noncompact T duality in $\phi$. Indeed after a timelike T duality the first line of eqn. \eqref{eq:mirrormetric} becomes
\begin{equation}
-(1-r^2)dt^2 + \frac{dr^2}{1-r^2} + r^2 d\Omega_{3},
\nonumber
\end{equation}
which is $\mathrm{dS}_5$ in static coordinates. Similarly, T duality in $\phi$ turns the second line into
\begin{equation}
(1+\rho^2) d\phi^2  + \frac{d\rho^2}{1+\rho^2} + \rho^2 d\Omega_{3},
\nonumber
\end{equation}
which is $\mathrm{H}^5$ in analogous coordinates. The timelike and noncompact nature of these T dualities makes this a rather formal relation however. So while our original string lived on the maximally symmetric space $\ads$, its mirrored version does not quite live on the maximally symmetric space $\mathrm{dS}_5 \times \mathrm{H}^5$, but rather its doubly T dual cousin \eqref{eq:mirrormetric}.

At this stage we should mention that based on simpler models \cite{Delduc:2013fga} it was conjectured that the maximal deformation limit of the models of the previous section should directly correspond to $\mathrm{dS}_5 \times \mathrm{H}^5$ \cite{Delduc:2013qra}. Indeed, this space was already extracted in a different $\varkappa \rightarrow \infty$ limit combined with two spacelike T dualities \cite{Hoare:2014pna}. This limit requires taking one of the coordinates outside its natural range however, thereby changing the coordinate that is timelike, and does not appear to be smoothly connected to the general deformed geometry. Like $\mathrm{dS}_5 \times \mathrm{H}^5$, the resulting geometry is formally supported by an imaginary five-form flux. Our limit instead yields a geometry that is smoothly related to the generic case, and naturally results in a real worldsheet theory. We consider our doubly T dual relation to $\mathrm{dS}_5 \times \mathrm{H}^5$, albeit part timelike and noncompact, in line with the conjecture of \cite{Delduc:2013qra}. Our mirror space is clearly related to the one obtained in \cite{Hoare:2014pna} by one time and three spacelike T dualities.

Our mirror space is a product of two five-dimensional spaces, with curvature
\begin{equation}
\nonumber
R = 4 \frac{1-2r^2}{1-r^2}-4\frac{1+2\rho^2}{1+\rho^2},
\end{equation}
showing a (naked) singularity at $r=1$. This means that in strong contrast to $\ads$, the mirror space is singular. Sigma models on singular backgrounds are not necessarily ill-defined however, and the integrability of our string sigma model is actually a promising indication of good behavior. 

A more pressing question is that of consistency of our mirror space as a string background, and we would like to show that it is part of a solution of supergravity. As our metric is related to the metric of $\mathrm{dS}_5 \times \mathrm{H}^5$ by two T dualities, it is natural to assume that our (type IIB) supergravity solution is supported by only a dilaton $\Phi$ and a self-dual five-form flux $F$, just as $\mathrm{dS}_5 \times \mathrm{H}^5$ formally is \cite{Hull:1998vg}. We could in fact induce our solution from the $\mathrm{dS}_5 \times \mathrm{H}^5$ one through the T dualities, but let us not assume familiarity with exotic supergravities. We then have to solve the equations of motion for the dilaton
\begin{equation}
\nonumber
4 \nabla^2 \Phi - 4(\nabla \Phi)^2 = R,
\end{equation}
the metric
\begin{equation}
\nonumber
R_{\mu\nu} = - 2 \nabla_\mu \nabla_\nu \Phi
+\frac{1}{4\cdot 4!}e^{2\Phi}F_{\mu\rho\lambda\sigma\delta}F_{\nu}^{\rho\lambda\sigma\delta} ,
\end{equation}
and the five-form
\begin{equation}
\nonumber
\pa_{\nu}\Big( \sqrt{-g} F^{\rho\lambda\sigma\delta\nu}\Big)=0 .
\end{equation}
The equation for the dilaton is solved by
\begin{equation}\nonumber
\Phi=\Phi_0-\frac{1}{2}\log  (1-r^2)(1+\rho^2),
\end{equation}
where $\Phi_0$ is a constant. Although the metric \eqref{eq:mirrormetric} corresponds to a direct product of two manifolds, the five-form has mixed components, $t$ matching up with the four transverse coordinates of the submanifold containing $\phi$ and vice versa \footnote{This corresponds to the exchange of the differentials $dt\leftrightarrow d\phi$ under the double T duality applied to the imaginary five-form that supports $\mathrm{dS}_5 \times \mathrm{H}^5$.}. Introducing would-be volume forms $\omega_t$ and $\omega_\phi$ for these two sets of five coordinates containing $t$ and $\phi$ respectively, the five-form is given by
\begin{equation}\nonumber
 F =  4 e^{-\Phi} \left( \omega_\phi - \omega_t \right).
\end{equation}

With this solution we can start adding Green-Schwarz fermions to the worldsheet theory in a canonical fashion. As indicated in the introduction, we can match these with an appropriate analytic continuation of the $\ads$ fermions at least at the quadratic level \cite{Arutyunov:2014jfa}. While the unboundedness of the dilaton raises questions regarding the mirror background in interacting string theory, this is no immediate problem for the sigma model of a free string.

This background has no conventional supersymmetry as it does not admit Killing spinors. This follows from the variation of the dilatino $\lambda$
\bea
\nonumber
\delta_{\eps}\lambda&=&\pa_{\mu}\Phi\, \Gamma^{\mu}\eps,
\eea
which does not vanish for any nonzero chiral spinor $\eps$ with our dilaton. Nevertheless, the string sigma model on our mirror background has a hidden form of supersymmetry \footnote{Still realized entirely within the scope of our free string, our hidden supersymmetry is not of the type discussed in \cite{Duff:1997qz}.}, inherited from the $\ads$ string.

\section{(Super)symmetry of the mirror model}

Since double Wick rotations preserve symmetries, our model must have the manifest $\mathfrak{psu}(2|2)^{\oplus2}$ symmetry of the light cone $\ads$ string. This symmetry need not be linearly realized however, and in fact the action of the supercharges cannot be. We can understand this from the form of the relevant superalgebras.

The on shell symmetry algebra of the light cone $\ads$ string is $\mathfrak{psu}(2|2)^{\oplus2} \oplus \mathbb{H}$, where the central element $\mathbb{H}$ corresponds to the worldsheet Hamiltonian. Considering one of the two copies of $\mathfrak{psu}(2|2)$, the supercharges $Q$ and $Q^\dagger$ satisfy
\begin{align}
\nonumber
\{Q_{\alpha}^{\,\,a}, Q^{\dagger \beta}_b \} = \delta^a_b R_\alpha^{\,\,\beta} + \delta_\alpha^\beta L_b^{\,\,a} + \frac{1}{2} \delta^a_b \delta_\alpha^\beta \mathbb{H},
\end{align}
where $L$ and $R$ generate the two bosonic $\mathfrak{su}(2)$'s. We see that as usual the generators of the superisometries of the light cone $\ads$ string anticommute to the generators of isometries, in particular time translations \footnote{In our light cone gauge the worldsheet Hamiltonian generates background time translations as well. Alternatively, this superalgebra looks completely natural from a hypothetical NSR point of view.}. For the mirror theory we instead have \cite{Arutyunov:2007tc}
\begin{align}
\nonumber
\{\tilde{Q}_{\alpha}^{\,\,a}, \tilde{Q}^{\dagger \beta}_b \} = \delta^a_b R_\alpha^{\,\,\beta} + \delta_\alpha^\beta L_b^{\,\,a} + T \delta^a_b \delta_\alpha^\beta \sinh{\frac{\tilde{\mathbb{H}}}{2}},
\end{align}
where the rest of the anticommutators vanish when we interpret the mirror theory as an on shell string with zero worldsheet momentum. Now were these mirror supercharges to correspond to linearly realized superisometries of a string background, they ought to anticommute to the associated Hamiltonian, not its hyperbolic sine. Put differently, identifying the mirror Hamiltonian as the generator of time translations, we are simply no longer dealing with a Lie superalgebra. Moreover, in the deformed models the symmetry algebra of the worldsheet theory is expected to be a quantum deformed version of $\mathfrak{psu}(2|2)^{\oplus 2}$, as supported by the S-matrix computations of \cite{Arutyunov:2013ega}. This symmetry does not appear to be realized geometrically on the deformed background however, and there is no reason to assume that it should be linearly realized on the worldsheet fermions, even in the maximally deformed limit where the algebra is no longer deformed. Note that the bosonic symmetry is nonetheless realized linearly, cf. the two three-spheres in eqn. \eqref{eq:mirrormetric}.

We should also briefly address the on and off shell symmetry algebras of the mirror theory. It is well known that the worldsheet symmetry algebra of the $\ads$ string picks up a central extension $\mathbb{C} \sim \sin \frac{\mathbb{P}}{2}$ when going off shell \cite{Arutyunov:2006ak}, allowing the exact S-matrix to be determined. In a string-based mirror theory, the situation should be as follows. We start with $\mathfrak{psu}(2|2)^{\oplus 2} \oplus \sinh\frac{\tilde{\mathbb{H}}}{2}$ worldsheet symmetry, where we can match $\sinh\frac{\tilde{\mathbb{H}}}{2}$ with $\mathbb{C}$ through analytic continuation. Going off shell by relaxing the level matching condition, the algebra should pick up a central extension matching the analytic continuation of $\mathbb{H}$. In this way the off shell symmetry algebras of both theories, and hence their exact S-matrices, would be related by the expected analytic continuation.

\section{Outlook}

We have introduced a transformation of background metrics that allows us to interpret mirror versions of light cone strings as light cone strings on different backgrounds. We then applied this to $\ads$ and gave the supergravity background in which a free string has a light cone worldsheet theory identical to the $\ads$ mirror theory. With this direct interpretation of the mirror model we can ask new types of questions, the main one being whether we can give meaning to this geometric mirror transformation in the context of AdS/CFT. The fact that a double Wick rotation on the worldsheet has an interpretation in terms of free string theory leads us to wonder whether a `similar analytic continuation' can be implemented in planar $\mathcal{N}=4$ SYM. If possible, the end result should have an interesting relation to our string. It might be fruitful to approach this through the deformed sigma models which continuously connect $\ads$ and the mirror background, where it may be possible to implement the deformation (perturbatively) in planar $\mathcal{N}=4$ SYM. The unbounded dilaton in our mirror background does not bode well for attempts at extending this beyond the sigma model of a free string (planar gauge theory), but then a double Wick rotation loses its simple physical interpretation on a higher genus Riemann surface to begin with.

Our considerations clearly apply to many spaces other than $\ads$, though they are of course mainly of interest when a mirror model comes into play \footnote{This trick may generate backgrounds with hidden supersymmetry regardless of direct interest in a mirror theory.}. Our procedure for example directly applies to ${\rm AdS}_2\times {\rm S}^2 \times {\rm T}^6$, and ${\rm AdS}_3\times {\rm S}^3 \times {\rm M}^4$ supported by pure Ramond-Ramond fluxes. More interesting are ${\rm AdS}_4 \times \mathbb{CP}^3$ (see e.g. \cite{Klose:2010ki}), and ${\rm AdS}_3\times {\rm S}^3 \times {\rm M}^4$ supported by mixed fluxes \cite{Cagnazzo:2012se}, since its metric  and $B$ field respectively do not fit the simplifying assumptions of this letter.
We will relax these assumptions when discussing the worldsheet fermions \cite{Arutyunov:2014jfa}.

Returning to our mirror background, it would be interesting to investigate possible consequences of its singular nature, starting for example with an analysis of classical string motion in our mirror space. Also, extracting the explicit fermionic couplings in the deformed coset sigma models is a complicated but relevant problem which may simplify our limit. Finally, while we now see that the deformed geometry interpolates between two solutions of supergravity, demonstrating that it is one at any $\varkappa$ remains an interesting open question.

\begin{acknowledgments}
We would like to thank R. Borsato, S. Frolov, B. Hoare, G. Korchemsky, M. de Leeuw, J. Maldacena, T. Matsumoto, A. Tseytlin, S. Vandoren, and B. de Wit for discussions. S.T. is supported by the Einstein Foundation Berlin in the framework of the research project "Gravitation and High Energy Physics" and acknowledges further support from the People Programme (Marie Curie Actions) of the European Union's Seventh Framework Programme FP7/2007-2013/ under REA Grant Agreement No 317089. G.A. acknowledges support by the Netherlands Organization for Scientific Research (NWO) under the VICI grant 680-47-602. The work by G.A. is also a part of the ERC Advanced grant research programme No. 246974,  {\it ``Supersymmetry: a window to non-perturbative physics"} and of the D-ITP consortium, a program of the NWO that is funded by the Dutch Ministry of Education, Culture and Science (OCW).
\end{acknowledgments}

\bibliographystyle{apsrev4-1}
\bibliography{Stijnsbibfile}

\end{document}